\documentclass[fleqn,10pt]{article}

\usepackage[margin=2cm]{geometry}
\usepackage{moreverb,url}
\usepackage{natbib}
\usepackage{bm}
\usepackage{authblk}
\providecommand{\keywords}[1]
{
  \small	
  \textbf{\textit{Keywords---}} #1
}

\def\I{\mathbf{I}}
\def\A{\mathbf{A}}

\def \vect {\textrm{vec}}

\def\x{\mathbf{x}}

\def\L{\mathbf{L}}
\def\Y{\mathbf{Y}}
\def\y{\mathbf{y}}
\def\k{\mathbf{k}}
\def\Gam{\bm{\Gamma}}

\usepackage{mathtools, nccmath}
\usepackage{algorithmic}
\usepackage{algorithm}

\usepackage{amsmath}

\usepackage{amsfonts}       
\usepackage{nicefrac}       
\usepackage{amsfonts,amssymb}
\usepackage{amsthm}
\usepackage{booktabs}       
\usepackage{microtype}
\usepackage{bm}
\usepackage{subfig}
\usepackage{tabularx}
\usepackage{array}
\usepackage{soul}
\usepackage{caption}
\usepackage[
    colorlinks,
    bookmarksopen,
    bookmarksnumbered,
    citecolor=black,
    linkcolor=black,
    urlcolor=black,
    pdfencoding=auto,
    psdextra
]{hyperref}
\def\I{\mathbf{I}}
\def\A{\mathbf{A}}

\def\x{\mathbf{x}}

\def\F{\mathbf{F}}

\def\L{\mathbf{L}}
\def\Y{\mathbf{Y}}
\def\k{\mathbf{k}}

\title{Spatial Thickness Mapping in Heterogeneous Plate Using Wave Physics-Informed Regression}

\author[3]{Amanda Beck}
\author[1]{Harsha Vardhan Tetali}
\author[2]{Michael MacIsaac}
\author[1]{Charlie Tran}
\author[1]{Woohyun Eum}
\author[2]{Ghatu Subhash}
\author[1]{Joel B. Harley}
\affil[1]{Department of Electrical and Computer Engineering, University of Florida, Gainesville, FL 32611, USA}
\affil[2]{Department of Mechanical and Aerospace Engineering, University of Florida, Gainesville, FL 32611, USA}
\affil[3]{Department of Physics, University of Florida, Gainesville, FL 32611, USA}

\begin{document}

\maketitle

\begin{abstract}
Traditional guided wave methods for structural health monitoring typically assume uniform material properties, which limit their ability to characterize heterogeneous structures with spatially varying thickness, damage, or material properties. These are challenges commonly encountered in corrosion assessment, composite delamination detection, and structural degradation monitoring. This paper presents a wave physics-informed regression approach that enables spatially resolved characterization of material properties by extracting local dispersion curves across a structure. Our approach focuses on a highly interpretable but flexible physics-informed framework that can be solved using fast algorithms and achieve robust numerical solutions. This paper discusses the mathematical design of the framework, the algorithm, and its interpretation. The framework was applied to a guided wave wavefield imaging dataset from a thin aluminum plate with non-uniform thickness around a hole to validate its practicality. The framework creates an accurate thickness map (correlation coefficient 0.94 with x-ray CT validation) as well as extracts the frequency-dependent velocities of waves within those regions.
\end{abstract}


\section{Introduction}
\label{sec: intro}

Guided ultrasonic waves have been widely applied to structural health monitoring. These waves are guided by the geometry of their propagation medium and are a widely researched tool for detecting, locating, and characterizing damage in physical structures \cite{Cawley2003-xm,Cawley2007-bq,Cawley2018-yd,kundu_mechanics_2019}. They can interrogate large areas with relatively low attenuation and have been implemented in a large variety of structural systems, including pipelines  \cite{Cawley2007-bq,Davies2009-nv,Liu2012-qi,Ying2010-sm,Ying2013-tz,Rizzo2005-lg,Li2002-xm}, bridges \cite{Chen2013-os,Holford2001-mm}, concrete \cite{Chong2003-yn}, steel cables \cite{Baltazar2010-na,Rizzo2004-sj,Rizzo2001-ot}, metal aircraft components \cite{Ihn2008-ai,Worden2003-pd}, and composite aircraft components \cite{Leckey2014-dk,Staszewski2009-pq,Qiu2011-cc,Salamone2009-zr}. Guided waves are often measured with densely sampled wavefield imaging systems, such as a scanning laser Doppler vibrometer \cite{CASTELLINI20061265}. Material flaws are then identified from patterns in the measured wavefield \cite{staszewski2004structural, mallet2004structural, leong2005structural}. However, these methods typically assume uniform material properties, limiting their effectiveness in heterogeneous or damaged structures.

In this investigation, we focus on Lamb waves, which travel through the thickness of their guide \cite{kundu_mechanics_2019}. Physically, Lamb waves are governed by a collection of dispersion curves that describe how the wavenumber (or equivalently, phase velocity) varies as a function of frequency. Each dispersion curve represents a different possible solution to the wave equation, which is commonly referred to as a mode. The full wavefield is a linear combination of modes. Dispersion relations for Lamb waves are governed by the Rayleigh-Lamb equations \cite{kundu_mechanics_2019}. In this representation, the modes are usually separated into symmetric and antisymmetric components (across the thickness axis).  

\begin{figure}[h!]
    \centering
    \includegraphics[width=0.9\linewidth]{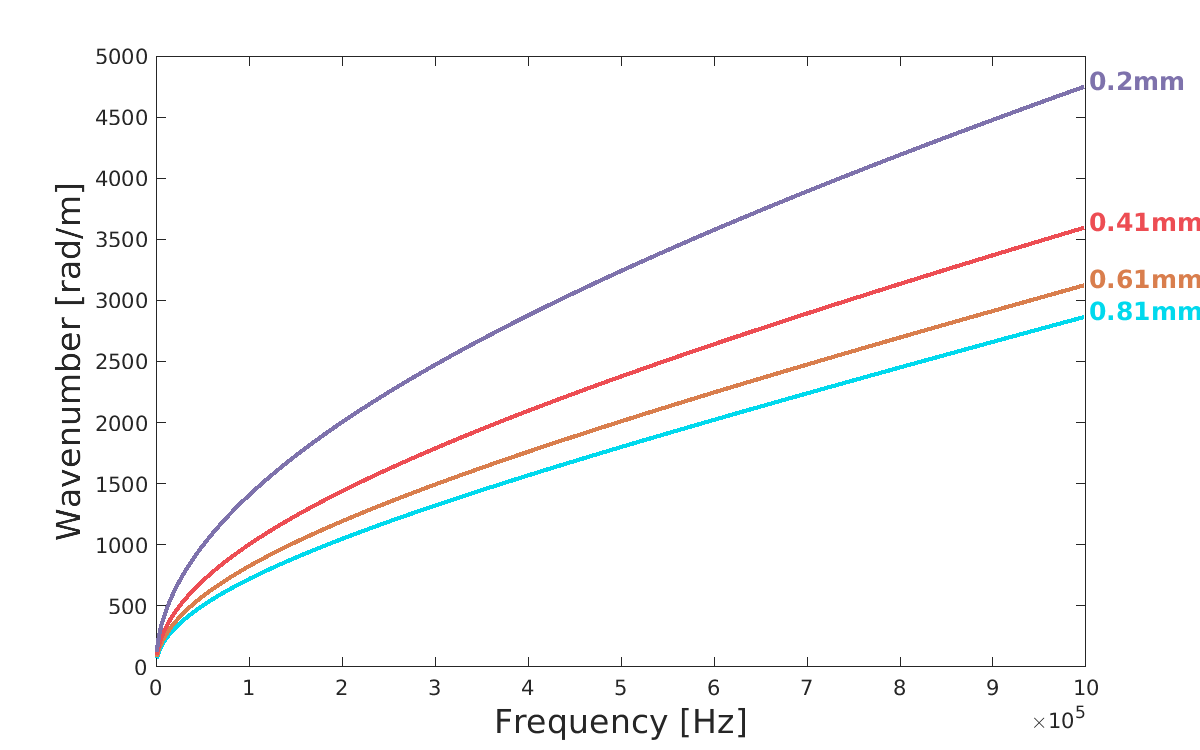}
    \caption{Example A0 dispersion curves for aluminum plates of varying thicknesses. These curves show how a variation in thickness changes the dispersion relation. The corresponding thicknesses are denoted beside each curve.\label{fig: dispcurves}}
\end{figure}

As material properties (such as anisotropy \cite{tran2025,subhash2025} and thickness) change, so do the dispersion curves. The Rayleigh-Lamb equations are complex, so determining how dispersion curves change with material properties can be done by numerically computing solutions \cite{zhu_effective_2025,wangensteen_pipe_2024}. An example set of first antisymmetric (A0) dispersion curves for aluminum plates of different thicknesses is shown in Fig.~\ref{fig: dispcurves}. For the A0 mode specifically, wavenumber and thickness are inversely related. Dispersion curves can be used to build predictive guided wave models \cite{subhash2025, sabeti2019sparse} to improve flaw detection, localization, and characterization methods through comparison with the model \cite{Leckey2014-dk}. They can also be used directly for diagnostics. For example, the local wavefield velocity or wavenumber can be used to identify delaminations \cite{app8091600} and porosity in composites \cite{Reed1993PorosityMI}. Dispersion curves can also be used to image the thickness loss in a metallic structure \cite{ruikunNTCEMC}.

However, a critical limitation in guided wave structural health monitoring is that most existing dispersion curve extraction methods assume uniform material properties throughout the structure. For example, traditional Fourier transform-based approaches \cite{Alleyne1990-yz} and compressive sensing methods for dispersion curve extraction, like sparse wavenumber analysis \cite{harley2013sparse}, mathematically assume the same velocities with the same amplitudes throughout a structure. This assumption fails in practical scenarios involving localized damage, material degradation, or inherently heterogeneous structures. Local wavefield methods have considered spatial windows to extract local dispersion information \cite{FLYNN20131, tian_rapid_2016, mesnil_instantaneous_2015, fan_lamb_2018, JEON2020102343, FLYNN20131}. Yet, since these methods independently extract dispersion curves around local regions, they will have limited wavenumber resolution, could be vulnerable to noise or wave scattering, and cannot easily segment regions with different material properties. 

To address these SHM challenges, this work presents a highly interpretable wave physics-informed regression approach that enables spatially resolved material characterization in heterogeneous structures. Our method is grounded in wave physics and requires no training data. While the deep learning literature has physics-informed \cite{karpatne2017theory} or physics-guided techniques, these approaches are known for being computationally difficult to train and sensitive to local sub-optimal solutions. Hence, we instead focus on a shallow, interpretable framework that utilizes similar regularization principles as those in deep physics-informed learning  \cite{tetali19wave,tetali2021wave,tetali2022physics, harley2022unsupervised, tetali2023learning, damiano2024zero}. This streamlined approach provides a nearly convex algorithm \cite{WIMFTSP} and a fast computational solution (due to the ability to operate in a sparse spectral domain). Regions with similar dispersion characteristics in the heterogeneous structures are then treated like spatially varying wave modes at each frequency.

In this paper, we derive wave-informed regression and demonstrate the algorithm with experimental ultrasonic wavefield data from a specimen scanned with a laser Doppler vibrometer. Specifically, we validate this approach through thickness mapping of a defective aluminum specimen, achieving accurate spatial characterization of material variations with 0.94 correlation to x-ray CT ground truth. While demonstrated through thickness mapping, this approach applies to various forms of structural heterogeneity, including corrosion, delamination, and material property variations. This demonstrates the method's capability for practical structural health monitoring applications where traditional uniform-property assumptions fail. We derive the wave-informed regression framework and demonstrate how results can be interpreted across frequency, wavenumber, and spatial domains.

\section{Wave-informed Regression Framework}
\label{sec: math}

For the reference of the readers, Table~\ref{tbl: notation}  lists the key variables and notation used in this paper.

\begin{table}[h]
\centering
\renewcommand{\arraystretch}{1.3}
\caption{List of key variables used in the paper. \label{tbl: notation}}
\begin{tabular}{| m{0.65cm} || m{12cm}|}
 \hline
 $x,y$ & Spatial coordinates.\\

 $\mathbf{X}$ & Matrix form of the measured wavefield data (2~spatial dimensions).\\
 
 $\mathbf{Y}$ & Matrix form of the learned solution to the Helmholtz equation (2~spatial dimensions). \\

 $\mathbf{x}$ & Vectorized form of the data  ($\vect(\mathbf{X})=\mathbf{x}$).\\
 
 $\mathbf{y}$ & Vectorized  learned solution \mbox{($\vect(\mathbf{Y})=\mathbf{y}$).}\\

  $k$ & Learned wavenumber \\

 $\mathbf{P}(k)$ & Physics operator corresponding to the Helmholtz equation with parameter $k$ \\

  $\gamma^2$ & Learned regularization parameter \\

  $\widehat{\;}$ & Hats over variables indicate data is in the spectral (i.e., wavenumber) domain \\

  \hline 
\end{tabular}
\end{table}

\begin{figure*}[ht!]
    \centering
    \includegraphics[width=1\linewidth]{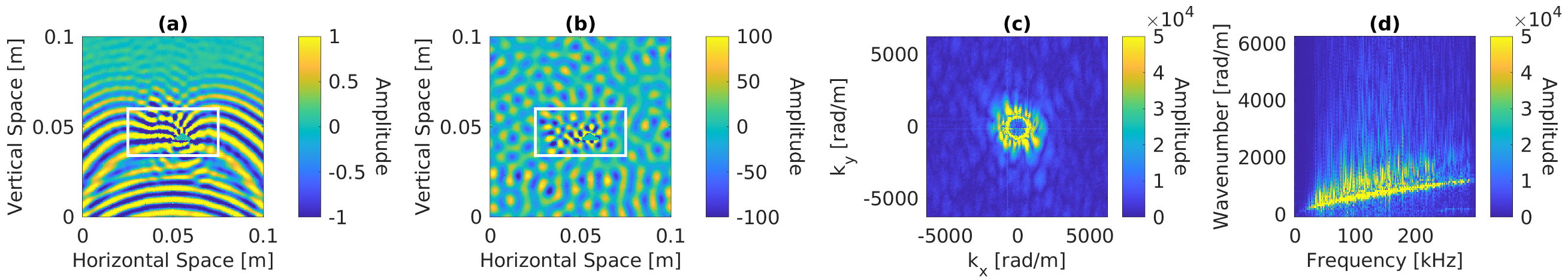}
    \caption{(a) Space-Time domain example wavefield used to illustrate wave-informed regression, at a single time step. (b) Real-valued, Frequency-Space domain wave-field from the data set in (a). (c) The spectral wavenumber representation of (b). Both (b) and (c) correspond to $\omega = 100 \ kHz$ (d) The Frequency-Wavenumber (dispersion) domain from (a).\label{fig: data}}
\end{figure*}

Given these variables,  we define a single-mode wave-informed regression by
\begin{align}
\label{eq: main_obj}
&\min_{\substack{\y, k, \gamma^2}} \left\| {\x} - {\y}  \right\|_2^2 + \gamma^2 \left\| {\y}  \right\|^2_2 + \frac{1}{\gamma^2} \left\| {\mathbf{P}(k)\y}  \right\|^2_2
\end{align}
where $\mathbf{x}$, $\mathbf{y}$, $\gamma^2$, and $\mathbf{P}(k)$ represent measured data, reconstructed solution, regression parameter, and physics operator, respectively. This optimization has three terms: the mean squared error cost, the energy cost, and the physics cost. The mean square error cost aims to have the solution $\y$ best match the data $\x$. The physics cost ensures the solution $\y$ satisfies or approximately satisfies the Helmholtz equation, represented by $\mathbf{P}(k)$ and more rigorously defined in the next subsection. The energy cost minimizes the size of the solution $\y$ and prevents $\gamma = \infty$ and $\mathbf{y} = \mathbf{x}$ from being the trivial optimal choices.  The variable $\gamma^2$ serves as a regression parameter but can also be interpreted as a bandwidth in the wavenumber domain, which we will demonstrate later. Notice that the optimization learns the solution $\y$ as well as the corresponding wavenumber $k$ and bandwidth $\gamma^2$ simultaneously. 

Ultimately, we are interested in simultaneously learning multiple wave modes within a dataset. Isolating individual wave modes would not only provide insight into material properties but wave-defect interactions could also create additional modes related to defect properties (e.g., thickness variations). This would result in multiple dispersion curves relating to the varying properties of the medium, as seen in Fig.~\ref{fig: dispcurves}. Therefore, we define a multi-mode wave-informed optimization framework by 
\begin{align}
\label{eq: cost_function}
    \min_{\substack{\y, \k, \bm{\gamma}^2}} 
    &\left\| {\x} - \sum_{m=1}^M{\y}_m  \right\|_2^2  \\
    &+ \sum_{m=1}^M \left[ \gamma_m^2 \; \left\| {\y_m}  \right\|^2_2 + \frac{1}{\gamma_m^2}  \left\| {\mathbf{P}_m(k_m) \y_m}  \right\|^2_2 \right] \nonumber
\end{align}
where $\mathbf{y}_m$ corresponds to a single mode solution to the Helmholtz equation. Furthermore, 
\begin{align}
    \y &= [\y_1, \ldots, \y_M] \\
    \k &= [k_1,\ldots, k_M]^T \\
    \bm{\gamma}^2 &= [\gamma_1^2, \ldots, \gamma_M^2]^T
\end{align}
correspond to the sets of all solutions, wavenumbers, and bandwidths that need to be learned. In this implementation, we assume the number of modes $M$ is known. 

Figure \ref{fig: data} shows the example data set used in this paper from four different perspectives. Figure~\ref{fig: data}(a) shows a time snapshot of a wave propagating through an aluminum plate with a thickness-varying defect around the center, indicated by the white box. Details on this plate are found in Section~\ref{sec: experimental-set-up}. Figure~\ref{fig: data}(b) shows the frequency domain representation of the same data set at $\omega = 100$~kHz. Figure~\ref{fig: data}(c) shows the 2D spectral (wavenumber) representation at $\omega = 100$~kHz, showing a circle representing the dominant wavenumber value with less distinct high wavenumber features corresponding to the thickness change. Relatedly, Fig.~\ref{fig: data}(d) shows the frequency-wavenumber domain, or dispersion curves, showing one strong A0 mode with additional higher wavenumber components corresponding to the thickness change. From the perspective of wave-informed regression, this data consists of a collection of effective modes that describe this spatially varying thickness. 

\subsection{Derivation of the Physics Operator} 
\label{sec: derivation_physics}

This subsection discusses how to derive the physics operator or matrix $\mathbf{P}(k)$ for the Helmholtz equation. The wave equation with two spatial dimensions is 
\begin{align}
    \frac{\partial^2 \mathbf{u}(t,x,y)}{\partial t^2} = c^2 \left( \frac{\partial^2 \mathbf{u}(t,x,y)}{\partial x^2} +  \frac{\partial^2 \mathbf{u}(t,x,y)}{\partial y^2} \right)
\end{align}
where $\mathbf{u}$ is a continuous wave function and $c$ is the phase speed of the propagating wave. By computing the Fourier transform of both sides of the equation, the time derivative can be eliminated, resulting in the simpler expression known as the Helmholtz equation. 
\begin{align}
\label{eq: 2dHelmholtz}
    k^2\mathbf{u}(\omega,x,y) = - \, \left( \frac{\partial^2 \mathbf{u}(\omega,x,y)}{\partial x^2} +  \frac{\partial^2 \mathbf{u}(\omega,x,y)}{\partial y^2} \right)
\end{align}
Notably, the Helmholtz equation is an eigenvalue equation for the Laplacian, where $k^2 = \omega^2 / c^2$ is both the wavenumber of the propagating wave and the eigenvalue. This way, solving the Helmholtz equation is equivalent to solving an eigenvalue problem. This investigation focuses on obtaining solutions to the Helmholtz equation at individual frequencies. Each frequency is independent of the others, so we may solve the wave equation at each frequency and then concatenate our results to recover the full spectrum wavefield. For a given scenario, there may be multiple solutions to the Helmholtz equation, each with a different $k$. Each solution is a mode of the overall multi-mode solution. 

Using matrix notation, the two-dimensional Helmholtz equation in \eqref{eq: 2dHelmholtz} can be expressed as
\begin{align}
    \label{eq: helmholtz_matrix}
    \Delta^2 k^2\mathbf{Y} &=  - \,  \left( \mathbf{Y} \L_x  +  \L_y \mathbf{Y}   \right) \\
    \label{eq: helmholtz_laplace}
    \Delta^2 k^2\mathbf{y} &=  - \, (\L_x \oplus \L_y) \mathbf{y} \,
\end{align}
where $\oplus$ represents a Kronecker sum. A proof of this relationship is shown in \ref{appendix: A}. In these expressions, $\mathbf{Y} \equiv \mathbf{Y}(\omega)$ and $\mathbf{y} \equiv \mathbf{y}(\omega)$ are the single-frequency representations for the propagating wave solution. In the case of wave-informed regression, this solution would be our reconstructed wavefield. The variables $\L_x$ and $\L_y$ represent the Laplacian (second derivative) operators/matrices in the $x$ and $y$ directions, respectively. We will discuss our definition of these matrices in Section~\ref{sec: spectral}. The term $\Delta$ is the wavenumber step size associated with the spatial sampling rate. 

The solution to our optimization framework must satisfy \eqref{eq: helmholtz_laplace} in order to follow the Helmholtz equation. By rearranging the variables in \eqref{eq: helmholtz_matrix}, we get that
\begin{align}
    \label{eq: helmholtz_laplace_physics}
    &\mathbf{P}(k) \mathbf{y}  = \mathbf{0}  \\
    \label{eq: helmholtz_laplace_physics2}
    &\mathbf{P}(k) = \Delta^2 k^2\I + (\L_x \oplus \L_y) \; . 
\end{align}
where $\mathbf{P}(k)$ is the physics operator in \eqref{eq: main_obj}. Hence, the optimal solution to the Helmholtz equation is one in which our solution $\y$ is contained within the null space of the Helmholtz operator $\mathbf{P}(k)$.

\subsection{Spectral Representation of the Physics Operator}
\label{sec: spectral}

Our algorithm can be computationally fast by performing the regression exclusively in the spectral domain, where our matrix multiplications in the cost function become element-wise. For the two-dimensional Helmholtz formulation, we can use Eigenvalue decomposition to modify \eqref{eq: helmholtz_laplace_physics} and \eqref{eq: helmholtz_laplace_physics2} to express the discretized wave equation as
\begin{align}
    \label{eq: helmholtz_laplace_physics3}
    &\widehat{\mathbf{P}}(k) \widehat{\mathbf{y}}  = \mathbf{0}  \\
    &\widehat{\mathbf{P}}(k) = \Delta^2 k^2\I + \, \left(  \mathbf{K}_x \oplus \mathbf{K}_y  \right) \; \\
    &\widehat{\y} = (\Gam_y \otimes \Gam_x) \mathbf{y} \nonumber \\
    &\L_x = \Gam_x \mathbf{K}_x \Gam_x^{-1} \nonumber \\
    &\L_y = \Gam_y \mathbf{K}_y \Gam_y^{-1} \nonumber
\end{align}
where $\mathbf{K}_x \oplus \mathbf{K}_y$ is now a diagonal matrix of eigenvalues $k^2$, and $\Gam_x$,  $\Gam_y$ are the eigenvectors of $\mathbf{L}_x$ and $\mathbf{L}_y$, respectively. A more detailed derivation can be found in \ref{appendix: B}. Essentially, this process is equivalent to solving the eigenvalue Helmholtz equation. Notably, in the special case of periodic boundary conditions, $\Gam_x,\ \Gam_y$ are simply the discrete Fourier transform operators, and so diagonalization can be achieved by performing a fast Fourier transform on the spatial dimensions in addition to the temporal dimension. A proof of the relationship between the Laplacian eigendecomposition and the Fourier transform can be found in~\ref{appendix: C}. For efficiency purposes, periodic boundary conditions are assumed in our implementation.

\subsection{Single Mode Optimization Solutions}
\label{sec: single_mode_sol}

While simultaneously optimizing all parameters in our single-mode framework is computationally challenging, each parameter has fast, closed-form solutions when all other parameters are assumed fixed. In this subsection, we will derive each of these solutions, which we use in our algorithmic implementation. Further details on these derivations can be found in \ref{appendix: D}.

\begin{figure}[ht!]
    \centering
    \includegraphics[width = .8\linewidth]{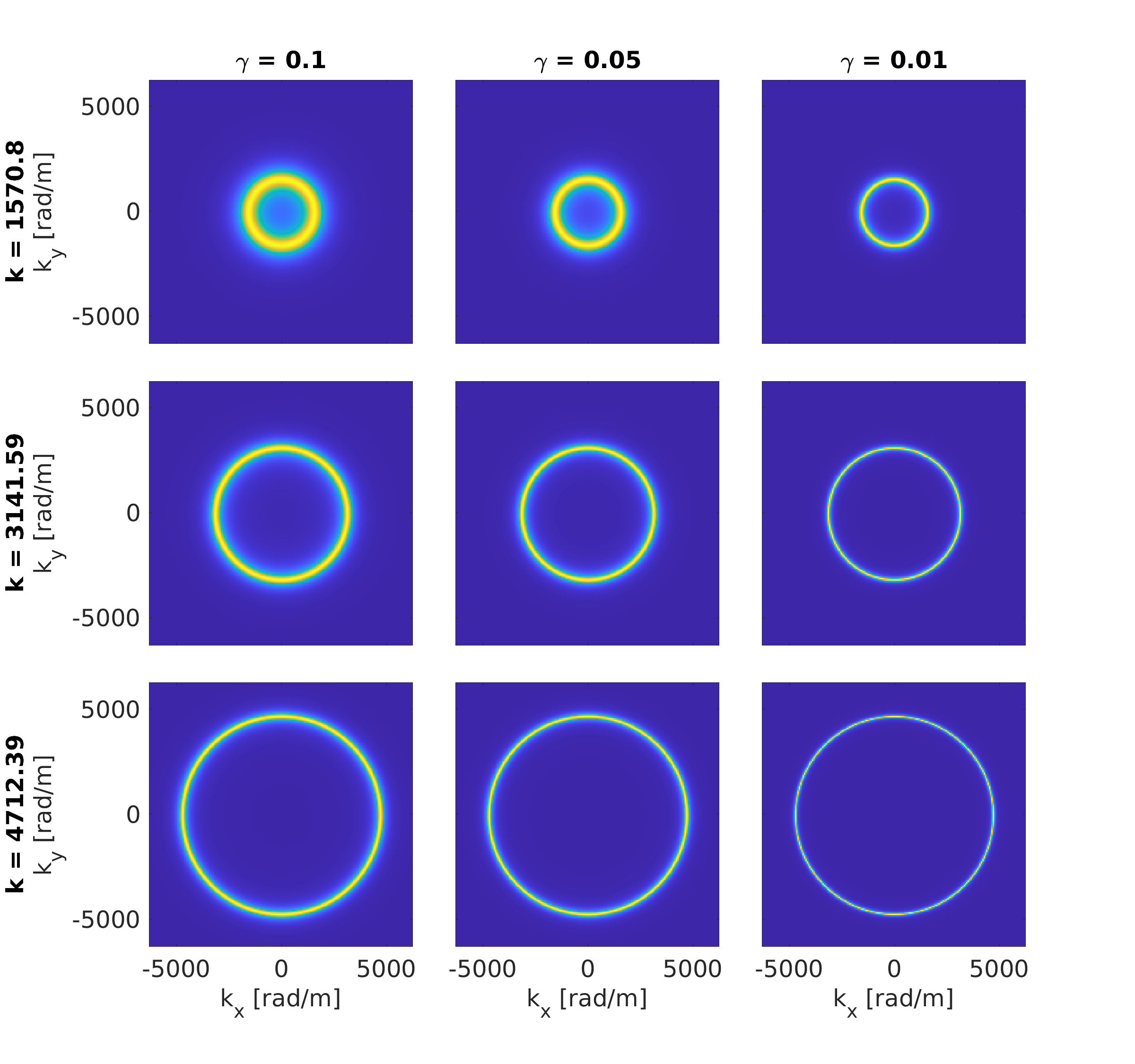}
    \caption{Example filters $\F(k)$ in the spectral wavenumber domain with varying wavenumbers $k$ and bandwidth $\gamma$. \label{fig: 2d_helmholtz_filter}}
\end{figure}

If we assume $k^2$ and $\gamma^2$ are known, we can learn the optimal solution $\y$ for our cost function in \eqref{eq: main_obj}. By computing the derivative of \eqref{eq: main_obj} with respect to $\y$ we can compute the derivative of our cost function and set it equal to zero. By solving for $\y$, we obtain the closed form solution  
\begin{align}
\label{eq: closed_form_y}
    \y &= \left( (1 + \gamma^2) \I  + \frac{1}{\gamma^{2}} \mathbf{P}(k)^H \mathbf{P}(k)  \right)^{-1} \x \\
    &= \F(k, \gamma) \x
\end{align}
The matrix $\F(k, \gamma)$ represents a mapping of our data onto a domain that approximately satisfies the physics. In the spectral representation of the physics operator discussed in Section~\ref{sec: spectral}, the diagonal $\F(k, \gamma)$ matrix acts as a band-pass filter around the exact physics solution. Figure~\ref{fig: 2d_helmholtz_filter} shows this band-pass behavior, as well as how the spectral domain filter changes as $k$ and $\gamma$ change. As $k$ increases, the radius of the filter increases. As $\gamma$ increases, the spread of the filter increases. Based on this observation, we may interpret $\gamma$ as an uncertainty parameter, or bandwidth. Effectively, $\gamma$ is the balancing factor between following the data and following the physics in our regression framework.

If we assume $\y$ and $\gamma^2$ are known, we can learn the optimal value of $k^2$ by similarly solving for the derivative with respect to $k^2$. We choose to optimize for $k^2$ instead of $k$ since $k^2$ are the direct eigenvalues of the differential equation. For our cost function, the optimal solution to $k^2$ is defined by 
\begin{align}
\label{eq: closed_form_k}
    k^2 = {\frac{\y^H \left[ \mathbf{P}(0) + \mathbf{P}(0)^H \right] \y}{ 2 \y^H \y }}
\end{align}
where $\mathbf{P}(0) = \L_x \oplus \L_y$ is the Laplacian term of the physics operator. This can be interpreted as the ratio of the inner product of the second derivative of $\y$ (across multiple dimensions) with the inner product of $\y$ with itself. We can see that in one dimension (as an example), when this function is applied to data that satisfies the Helmholtz equation (e.g., $\cos(k x)$), this equation yields the $k^2$. Referring back to Fig.~\ref{fig: 2d_helmholtz_filter}, $k$ is not only the estimated wavenumber, but also the radius of the spectral domain filter.

If we assume $\y$ and $k^2$ are known,  we can learn the optimal value of $\gamma^2$ by similarly solving for the derivative with respect to $\gamma^2$. For our cost function, the optimal solution to $\gamma^2$ is defined by
\begin{align}
\label{eq: closed_form_g}
    \gamma^2 &= \sqrt{\frac{ \y^H \mathbf{P}(k)^H  \mathbf{P}(k) \y }{ \y^H \y  }} \;.
\end{align}
Since $\mathbf{P}(k) \y$ should equal $0$ when the physics is matched, we again interpret this as a measure of uncertainty around the exact physical solution. 

\subsection{Multi-Mode Optimization Solutions}
\label{sec: multi_mode_sol}

If we assume $k_m^2$ and $\gamma_m^2$ are known for each mode, we can learn the optimal solutions $\y_m$ for our cost function in \eqref{eq: cost_function}. Solving for all modal solutions $\y_m$ simultaneously obtains the closed-form solution similar to \eqref{eq: closed_form_y}
\begin{align}
    [\mathbf{y}_1, \ldots, \mathbf{y}_M ] &= \left( \mathbf{A}^H \mathbf{A} +  \mathbf{G}  + \frac{1}{\mathbf{G}}\mathbf{P}(\mathbf{k})^H\mathbf{P}(\mathbf{k}) \right)^{-1} \mathbf{A}^H \x \\
     &= \F(\mathbf{k}, \bm{\gamma}) \x \nonumber
\end{align}
where our matrices for different modes are distributed across block matrices
\begin{align}
    \mathbf{P}(\mathbf{k}) &= 
    \begin{bmatrix}
        \mathbf{P}(k_1) &  &  & \\
        &  \ddots & \\
        &  & \mathbf{P}(k_M)
    \end{bmatrix}
\end{align}    
\begin{align}
    \mathbf{G} &=
    \begin{bmatrix}
        \gamma_1^2 \I &  &  & \\
        &  \ddots & \\
        &  &  \gamma_M^2 \I
    \end{bmatrix}
\end{align}  
that treat each mode as an independent element in the inversion process. The mixing matrix $\mathbf{A}$ consists of $M$ stacked identity matrices
\begin{align}
    \mathbf{A} &=
    \begin{bmatrix}
        \I & \I & \cdots & \I
    \end{bmatrix} \; .
\end{align}
that defines the relationship between modes (i.e., their sum should closely match $\x$).

From our cost function, each value of $k_m$ and $\gamma_m^2$ is in separate terms and therefore can be optimized with respect to one another. Therefore, the optimal solutions to $k_m^2$ and $\gamma^2_m$ are still defined by \eqref{eq: closed_form_k} and \eqref{eq: closed_form_g}, respectively. 

\subsubsection{Filter Bandwidth Interpretation}
\label{sec: bandwidth_variance}

In Figure~\ref{fig: 2d_helmholtz_filter}, $\gamma$ is proportionally related to the bandwidth of the spectral domain filter. However, $k$ also impacts it: as $k$ increases, the bandwidth decreases, despite $\gamma$ remaining constant, as is shown in each column of Fig.~\ref{fig: 2d_helmholtz_filter}. To evaluate the uncertainty, we compute the full width, half magnitude (FWHM) of the spectral, single-mode filter $\mathbf{F}(k, \gamma^2)$ at a single frequency. To compute this, assume the eigenvalues in diagonal matrices $\mathbf{K}_x$ and $\mathbf{K}_y$ are continuous values $-\kappa_x^2$, $-\kappa_y^2$. This is based on the fact that the Fourier transform of a second derivative of some spatial signal $f(x)$ is 
\begin{align*}
    \mathcal{F} \left\{ \frac{d^2 f(x)}{dx^2} \right\} = - \kappa_x^2 F(\kappa_x) \; ,
\end{align*}
where $F(\kappa_x)$ is the Fourier transform of $f(x)$.

Under this assumption, we can write the filter as
\begin{align}
    F(\kappa_x, \kappa_y, k, \gamma) = \frac{1}{1 + \gamma^2 + \frac{1}{\gamma^2} \left( \kappa_x^2 + \kappa_y^2 - k^2 \right)^2 }
\end{align}
The maximum value occurs when $\kappa_x^2 + \kappa_y^2 = k^2$ and the half-width amplitude would then be $1/(2(1 + \gamma^2))$. The filter $F(\kappa_x, \kappa_y, k, \gamma)$ has this amplitude when
\begin{align}
    \kappa_x^2 + \kappa_y^2 &= 
    k^2 \pm \gamma \sqrt{1 + \gamma^2} \; .
\end{align}
After taking the square root of both sides, the Taylor series expansion around $\gamma = 0$ reveals that 
\begin{align}
    \sqrt{\kappa_x^2 + \kappa_y^2} &= 
    k \pm \frac{\gamma}{2 k} + \mathcal{O}\left( \frac{\gamma^2}{k^3} \right)  \; .
\end{align}
The full width is twice the second term, so we get an approximate FWHM of 
\begin{align}
    \textrm{FWHM} \approx \gamma / k  \; .
\end{align}
Therefore, the bandwidth of the filter is approximately proportional to $\gamma$ and inversely proportional to $k$.

\section{Wave-Informed Regression Algorithm}
\label{sec: code}

\makeatletter
\newcommand{\UNNUMBEREDSTATE}[1]{%
  \let\@currentlabel\relax 
  \hspace{\algorithmicindent}#1}
\makeatother

\newcommand\algorithmicprocedure{\textbf{procedure}}
\newcommand{\algorithmicendprocedure}{\algorithmicend\ \algorithmicprocedure}
\makeatletter
\newcommand\PROCEDURE[3][default]{%
  \ALC@it
  \algorithmicprocedure\ \textsc{#2}(#3)%
  \ALC@com{#1}%
  \begin{ALC@prc}%
}
\newcommand\ENDPROCEDURE{%
  \end{ALC@prc}%
  \ifthenelse{\boolean{ALC@noend}}{}{%
    \ALC@it\algorithmicendprocedure
  }%
}
\newenvironment{ALC@prc}{\begin{ALC@g}}{\end{ALC@g}}
\makeatother

As discussed in Section~\ref{sec: math}, the optimization of the multi-mode cost function \eqref{eq: cost_function} in the spectral domain of the physics operator is essentially a filtering process. The final filter equation as well as the closed form solutions for the reconstruction $\mathbf{y}$ and optimization parameters $\gamma$ and $k^2$ were defined in \eqref{eq: closed_form_y}, \eqref{eq: closed_form_k}, and \eqref{eq: closed_form_g}, respectively. This section addresses how the regression is performed algorithmically. 

Algorithm \ref{algoblock: meta-algo} outlines the procedure. It is based on a coordinate descent process \cite{wright_coordinate_2015}, which iteratively solves for each variable, adjusting search ranges based on prior results, and reaches the optimal solutions based on a defined threshold. We initialize $\gamma_m^2=(2 \pi)^2$ since a bandwidth of $2 \pi$ covers the entire spectral domain. We initialize $k_m = 0$ since this represents setting the physics term to be $(\mathbf{L}_x \oplus \mathbf{L}_y )\mathbf{y} = 0$, essentially solving for the eigenvalues of the Laplacian, which are the $k_m^2$ values. Once the first solution $\mathbf{y}$ is computed, each term is then iteratively optimized through its respective closed form solution until the stopping threshold or the maximum number of iterations is met.

We can identify a stopping criterion after rewriting the cost function. Specifically, for the single-mode case, the cost function can be rewritten as 
\begin{align}
\label{eq: main_obj2}
&\min_{\substack{\y, k, \gamma^2}} \x^H \x - 2 \x^H \y +  \y^H \mathbf{F}^{-1}(k,\gamma) \y \; .
\end{align}
We observe here that $\y$ minimizes this cost function only when it also maximizes $\x^H \y$. Similarly, the multi-modal cost function
\begin{align}
\label{eq: main_obj3}
&\min_{\substack{\y, k, \gamma^2}} \x^H \x - 2 \x^H \A \y +  \y^H \mathbf{F}^{-1}(\k,\bm{\gamma}) \y \; .
\end{align}
shows $\y$ minimizes this cost function only when it also maximizes $\x^H \A \y = \x^H (\sum_{m=1}^M \y_m)$. Therefore, we choose a stopping criteria for which $\theta = \x^H (\sum_{m=1}^M \y_m)$ either converges or begins to decrease. In this implementation, we define convergence as a change of less than $\eta = 0.1$ or a maximum while-loop iteration count of $20$.

\begin{algorithm}[t!]
\caption{\bf{Coordinate Descent Optimization}}
\label{algoblock: meta-algo}
\begin{algorithmic}[1]
\PROCEDURE{coordinate-descent}{$\mathbf{x},M, \eta$}

    \STATE{$k_m^2 = 0$ \quad \textrm{for all } $m$ }
    \STATE{$\gamma_m^2 = (2\pi)^2$ \quad \textrm{for all } $m$ }
    \STATE{$\theta_0 = 0$}
    
    \WHILE{$\theta_n - \theta_{n-1} > \eta$}
    \FOR{$1 \leq m \leq M$}
        \STATE{$\gamma_m^2 = \sqrt{\frac{ \y^H \mathbf{P}(k)^H  \mathbf{P}(k) \y }{ \y^H \y  }}$}
        \STATE{$ [\mathbf{y}_1, \ldots, \mathbf{y}_M ] = \F(\mathbf{k}, \bm{\gamma}) \x$}
        \STATE{$k_m^2 = \frac{\y^H \left[ \mathbf{P}(0) + \mathbf{P}(0)^H \right] \y}{ \y^H \y }$}
        \STATE{$ [\mathbf{y}_1, \ldots, \mathbf{y}_M ] = \F(\mathbf{k}, \bm{\gamma}) \x$}
    \ENDFOR
    \STATE{$\theta_{n} = \mathbf{x}^H \left( \sum_{m=1}^M \mathbf{y}_m \right) $}
    \ENDWHILE
\ENDPROCEDURE
\end{algorithmic}
\end{algorithm}

\section{Experimental Setup}
\label{sec: experimental-set-up}

\begin{figure}[ht!]
    \centering
    \includegraphics[width=0.5\linewidth]{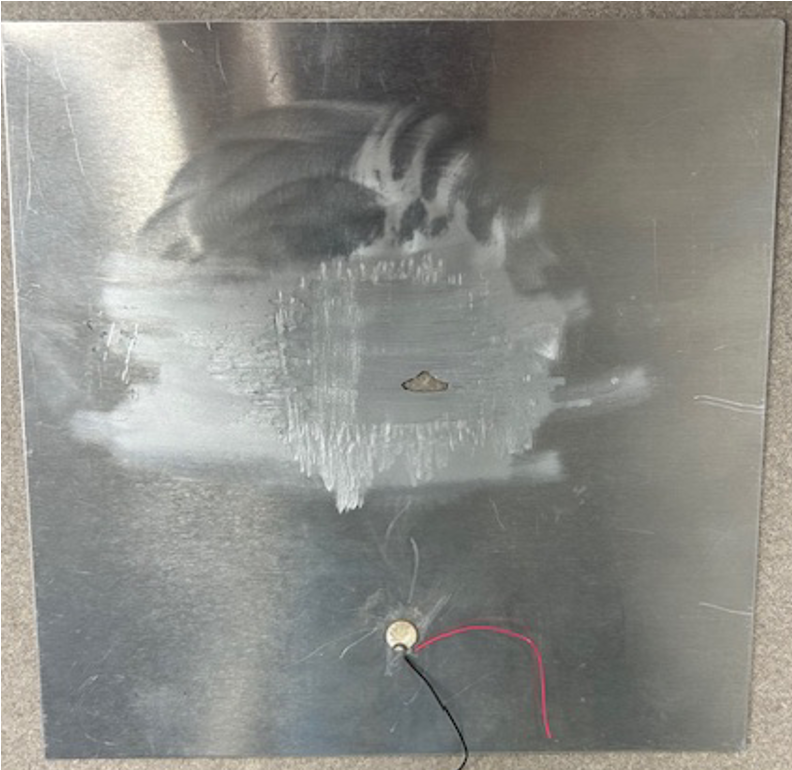} 
    \caption{Image of the plate specimen used for this experiment.}
    \label{fig: plate_image}
\end{figure}

We conducted an experiment on an aluminum plate of dimension $15.24 \times 15.24$~cm with thickness $0.8128$~mm, and a manufactured defect. The defect was created around the center of the plate by sanding down the specimen, creating a well with a hole at the bottom, as seen in Fig.~\ref{fig: plate_image}. The main objective of this experiment is to show that wave-informed regression can identify regions of varying thickness through mode decomposition. 

The ultrasonic wavefield was induced using a piezoelectric sensor fixed to the plate. A chirp signal \cite{MICHAELS2013265} of length 1000 with a maximum frequency of 300 kHz was emitted. The scan was performed using a laser Doppler vibrometer, and comprised an area of $10\times10$~cm around the defect. Note that we did not specially treat or polish the plate to improve reflectance and therefore improve signal-to-noise ratio. The number of spatial samples in $x$ and $y$ dimensions ($N_x$, $N_y$) is $200$. Thus, the spatial resolution ($d_x$) is $0.0005$~m. The number of time samples ($Q$) is $2000$. The temporal sampling rate ($F_s$) is $2$~MHz. 

The scan results in data as a function of $t$, $x$, and $y$. Before passing into our algorithm, the data is pulse compressed \cite{LAM1976111} and transformed into the frequency domain with the fast Fourier transform.

\section{Results \& Discussion}
\label{sec: results}

In the previous sections, it was established that wave-informed regression can theoretically separate multiple modes / wavenumber values for each frequency, allowing us to identify dispersion curves that correspond to various spatial regions with distinct characteristics. In this section, we illustrate the algorithm's performance with the experiment in Section~\ref{sec: experimental-set-up}. We look for $M = 4$ spatially separate wave modes that correspond to regions of distinct thicknesses. Our evaluation includes three main objectives: full wavefield reconstruction performance in Section~\ref{sec: rec_performance}, detection of spatial wavenumber variations in Section~\ref{sec: k_variations}, and thickness variation validation in Section~\ref{sec: thkns_validation}. The third objective consists of two tasks: estimate of thickness throughout the plate in Section~\ref{sec: thkns_validation}, and comparison of predicted wavenumbers to theoretical dispersion curves for ground truth thicknesses in Section~\ref{sec: disp_curves_estimates}. We lastly provide an interpretation of the uncertainty of our results in Section~\ref{sec: gamma}.

\begin{figure*}[h!]
    \centering
    \includegraphics[width=1\linewidth]{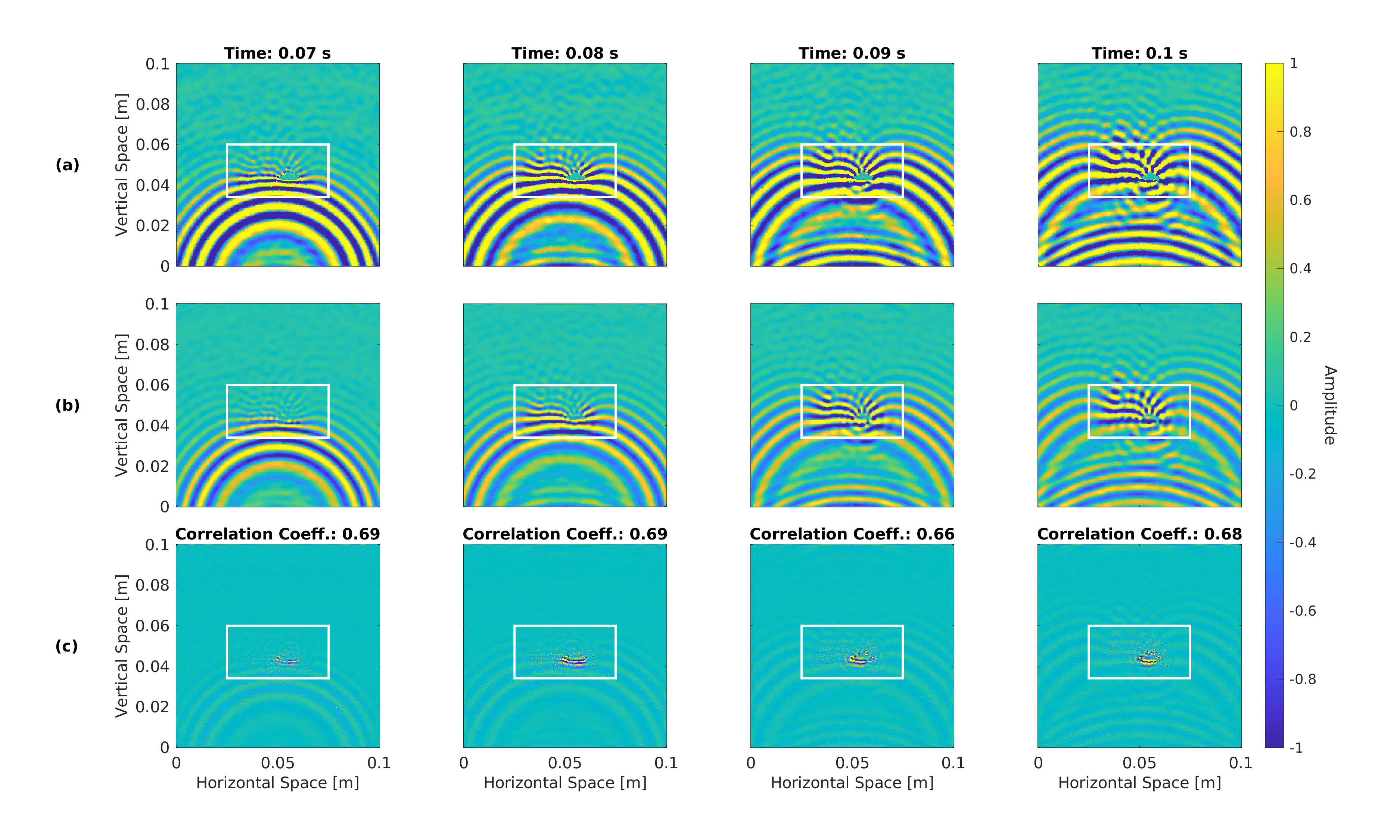} 
    \caption{(a) Depicts the original wavefield at 4 time steps. (b) Depicts the reconstructed wavefield at the same 4 time steps. (c) Depicts the residual error between the original and reconstruction. The associated correlation coefficients are displayed above each residual wavefield.\label{fig: dat_data_rec}}
\end{figure*}

\subsection{Objective 1: Reconstruction Performance}
\label{sec: rec_performance}

This objective evaluates how well wave-informed regression can reconstruct the original wavefield data. Figure~\ref{fig: dat_data_rec}(a)-(c) depicts the measured wavefield $\x$,
the reconstructed wavefield $\y$, and the corresponding residual $\mathbf{r}$ at four points in time. We compute the residual as
\begin{align}
    \mathbf{r} = \x \frac{\x^{T}\y}{\|\x\|_2^2} - \y \; ,
\end{align}
where the data at each mode $\x$ is multiplied by a scalar amplitude factor that minimizes the residual in the least squares sense. We use this normalization since we are primarily interested in the spatial and temporal wave propagation characteristics for defect detection. The amplitude difference is not significant. The residual wavefield shows that most error is located around the hole. This is expected since the hole represents a discontinuity in the wavefield that breaks the spatial continuity assumption of the Helmholtz equation. The overall correlation coefficient between the measured and reconstructed wavefield across all time is $0.68$. Much of the reduction from $1$ is due to complex signatures created by damage, as seen in the residual wavefield in Fig.~\ref{fig: dat_data_rec}(c). Visually, however, we observe a strong similarity with only 4 modes, and we observe that the error is primarily located around the defect regions. Hence, this can be used as a means of defect detection. 

\begin{figure*}[h!]
    \centering
    \includegraphics[width=1\linewidth]{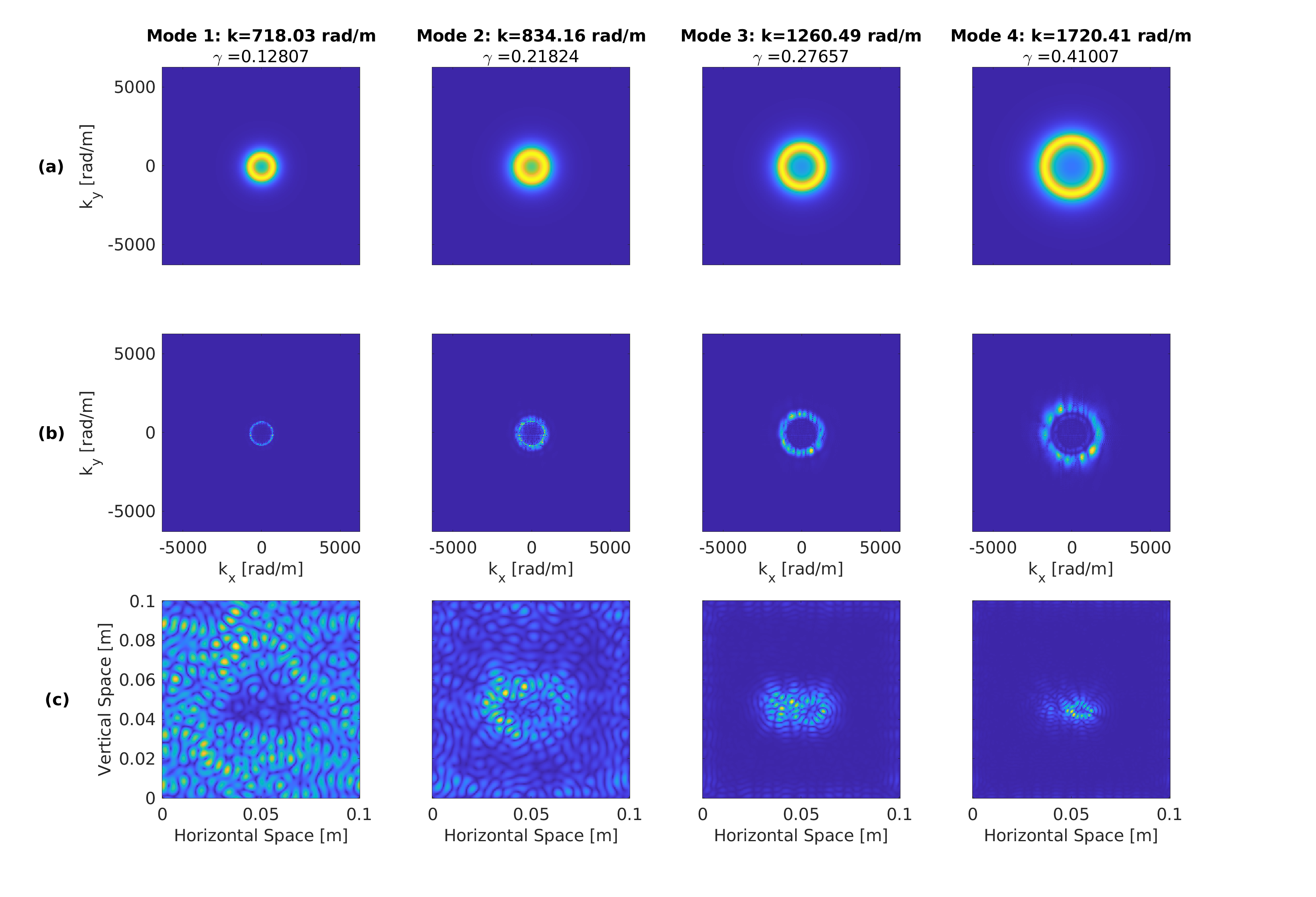}
    \caption{Each column corresponds to one of the 4 modes extracted. Row (a) depicts the optimal filter found for each of the 4 modes. Row (b) depicts the 4 extracted modes in the wavenumber-frequency domain. Row (c) depicts the 4 extracted modes in the spatial-frequency domain. All images correspond to a single temporal frequency of $100$~kHz. The optimal values found for $k_m$ and $\gamma_m$ are listed above each mode's column.\label{fig: dat_4_modes}}
\end{figure*}

\subsection{Objective 2: Detecting Spatial Wavenumber Variations}
\label{sec: k_variations}

This objective qualitatively evaluates whether wave-informed regression can spatially separate wave modes based on material changes by visually comparing mode results with Fig.~\ref{fig: plate_image}. Figure~\ref{fig: dat_4_modes} shows the estimated wavenumber domain filter, the frequency-space domain, and frequency-wavenumber domain of each extracted mode in rows (a)-(c), at a single frequency of $100$~kHz. Each column corresponds to a distinct mode. 

Mode~1, rows (a)-(b) show this mode has the smallest wavenumber $k_1 = 718.03$~rad/m. Mode 1, row (c) indicates that mode~1 primarily encompasses the non-defective, baseline area. A noticeable loss of amplitude is present around where the defect is located. The average thickness associated with mode~1 is, therefore, expected to be the highest. 
Mode~2, rows (a)-(b) show a slightly larger wavenumber, $k_2 = 834.12$~rad/m, than mode~1. Row (c) shows a concentration around the defective region, with some components in the overall plate. Thus, mode~2's associated average thickness is expected to be slightly lower than mode~1's. Note that modes~1 and 2 both fluctuate in their dominant regions as the frequency changes, so there is significant overlap in their overall results. 
Modes~3 and 4 rows (a)-(b) show increasing wavenumber values of $k_3 = 1260.49$~rad/m and $1720.41$~rad/m. Row (c) shows amplitude concentrations in regions of decreasing thickness. Mode 4 is dominant at the lowest thickness, and corresponds to the highest wavenumber. These results suggest that for the A0 mode, velocity decreases with thickness. Based on the spatial separability shown in Fig.~\ref{fig: dat_4_modes} and discussed above, we conclude wave-informed regression can separate baseline modes from defect-related modes. Our framework also identifies thickness variations within the defect. Finally, our wavenumber estimations display the expected inverse relationship with change in thickness, as predicted in Section~\ref{sec: intro} and Fig.~\ref{fig: dispcurves}.

\begin{figure*}[!t]
    \centering
    \includegraphics[width=1\linewidth]{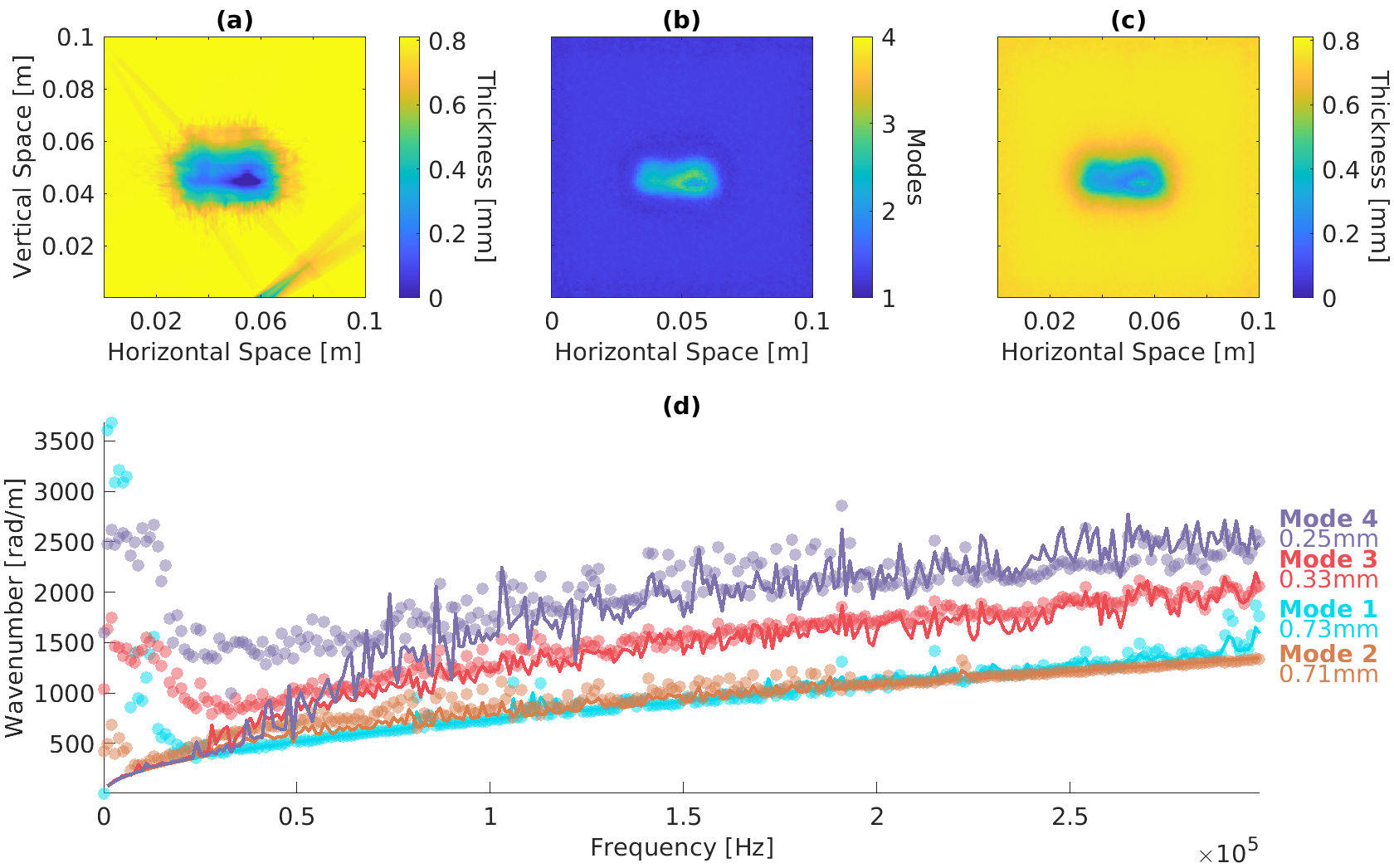}
    \caption{(a) The XCT scan of the specimen, integrated over the thickness axis. (b) Wave-informed regression mode amplitude distribution map averaged over all frequencies. (c) Reconstructed specimen thickness map from wave-informed regression dispersion curve estimations. (d) Wave-informed regression estimated (points) and XCT measured thickness theoretical (lines) dispersion curves for modes 1-4.\label{fig: thkns_map}}
\end{figure*}

\subsection{Objective 3: Thickness Validation}
\label{sec: thkns_validation}

This objective quantitatively evaluates whether wave-informed regression can map thickness variations through wavenumber estimation. We perform this validation in two ways: first, we assess whether the wavenumber estimates for each mode form dispersion curves that match the theoretical expectations. Second, we assess if we use this to build a map of thickness variations across the plate. 

For both assessments, the ground truth thickness map originates from an X-ray computed tomography scan, shown in Fig.~\ref{fig: thkns_map}(a). To compare the X-ray computed tomography scan with each frequency-dependent mode as shown in Fig.~\ref{fig: dat_4_modes}(c), we build a segmented image that maps each pixel to a single mode. The assignment is chosen by normalizing the magnitude of each frequency-dependent mode between 0 and 1 and then assigning the mode with the highest normalized magnitude to each pixel. We choose this normalization to compare modes in a common scale, thus avoiding favoring certain modes over others depending on the size of their associated area. Fig.~\ref{fig: thkns_map}(b) illustrates the average of these segmented maps across frequency. In general, we observe that each mode tends to be localized in certain regions, although the boundaries of these regions tend to contain a mixture of modes across frequency.

\subsubsection{Dispersion Curve Validation}
\label{sec: disp_curves_estimates}

Figure~\ref{fig: thkns_map}(d) compares the wave-informed regression wavenumber estimates (scatter plot) with the theoretical dispersion curves (solid lines). The solid lines represent the average wavenumber within each segmented mode region. We compute these theoretical wavenumbers by solving the Rayleigh-Lamb equation for the average observed X-ray computed tomography thickness in each region. 

We calculate the median residual error overall and for each mode to assess fit as $\epsilon = \textrm{median}(k_{\textrm{theory}} - k_{\textrm{WPIML}})$. The average error over all modes is $48.12$~rad/m. Modes 1-4 had errors of $18.81$~rad/m, $21.26$~rad/m, $40.83$~rad/m, and $101.21$~rad/m, respectively. Based on this analysis, our algorithm effectively estimated wavenumber values for each region's thickness, demonstrating the potential for defect characterization. 

\subsubsection{Thickness Mapping}
\label{sec: thkns_map}

Figure~\ref{fig: thkns_map}(c) illustrates a thickness mapping based on the wavenumbers associated with each frequency-dependent segmented mode region. We compute this thickness for each region by performing a coarse search (from a bank of 257 theoretical A0 dispersion curves for thicknesses ranging from $0$ to $0.8128$~mm) with the Rayleigh-Lamb equation to identify the thickness that matches the wavenumber learned for that mode. These thickness estimates are then averaged across frequency. The result is a reconstructed thickness map of the specimen. 

We exclude frequencies up to $50$~kHz in our averaged reconstruction due to the resolution of our measurements. 
Since the total length of our scan in $x$ and $y$ coordinates is $10$~cm, the maximum wavelength detectable is $10$~cm. Given a velocity of approximately $3000$~m/s, the associated minimum resolvable frequency would be $30$~kHz. In Figure~\ref{fig: thkns_map}(c), we see that our learned wavenumbers converge to the correct values around $30$~kHz. We exclude frequencies below $50$~kHz rather than $30$~kHz to reduce uncertainty associated with defining the correct cutoff.

The frequency-averaged thicknesses associated with modes 1-4 are $0.7235$~mm, $0.6586$~mm, $0.2621$~mm, and $0.1564$~mm, respectively. Overall, we observe a correlation coefficient of $0.94$ between the values of the X-ray computed tomography scan in Fig.~\ref{fig: thkns_map}(a) and our reconstructed map in Fig.~\ref{fig: thkns_map}(c), indicating a strong agreement in the shape of the images from both modalities. We therefore conclude that wave-informed regression can effectively reconstruct thickness variations.
 
\subsection{Uncertainty Interpretation}
\label{sec: gamma}

\begin{figure}[!t]
    \centering
    \includegraphics[width=1\linewidth]{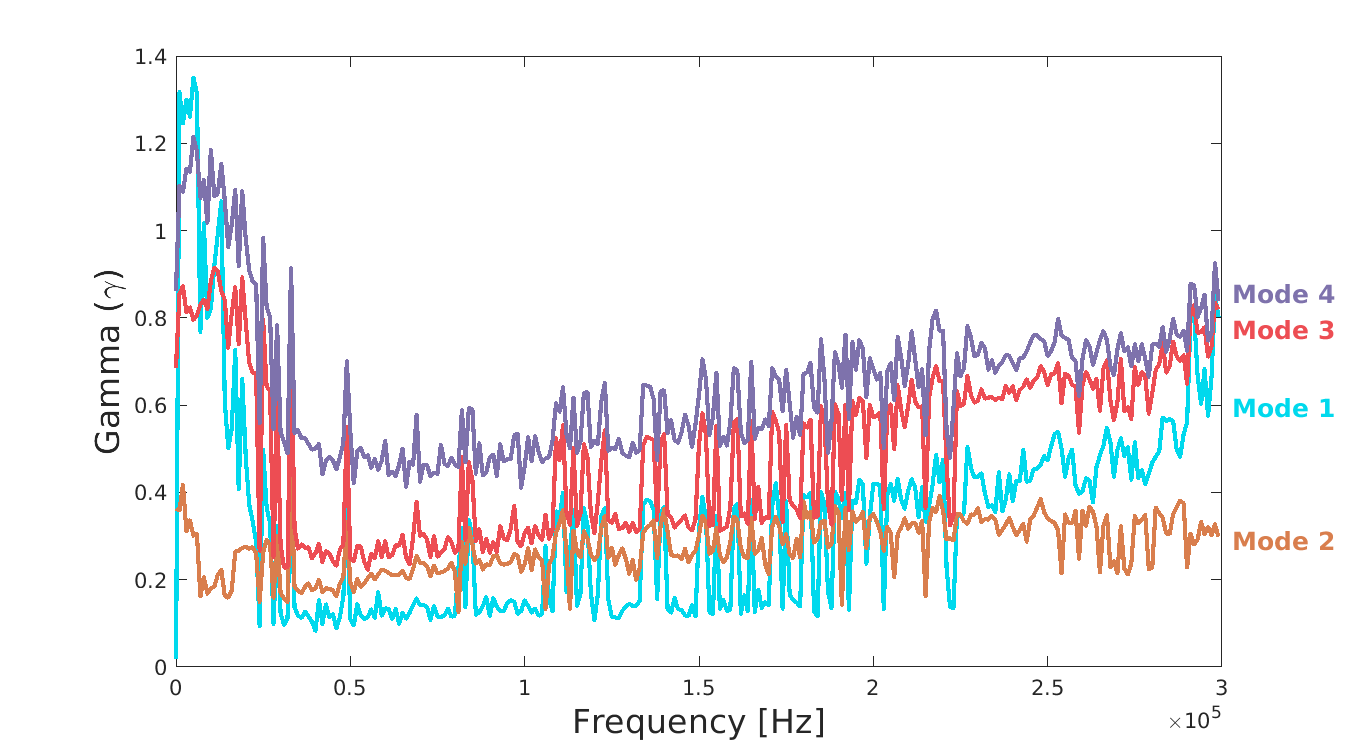}
    \caption{$\gamma$ (uncertainty) values for each mode. The magnitude of these values corresponds to the size of the spatial area covered by each mode. \label{fig: gamma}}
\end{figure}

Figure~\ref{fig: gamma} depicts the estimated $\gamma$ values for each mode at each frequency. We observe larger values in earlier frequencies, before approximately $30$~kHz, which aligns with our discussion on the detectable wavenumber range in Section~\ref{sec: thkns_map}. Past $30$~kHz, the $\gamma$ values overall increase for all modes. This trend is consistent with our discussion on the dependence of the filter bandwidth on both $\gamma$ and $k$ in Section~\ref{sec: bandwidth_variance}. As frequency increases, so does wavenumber. As wavenumber increases, the filter bandwidth tends to narrow, so, to account for uncertainty, $\gamma$ values increase.

The estimated $\gamma$ is also dependent on mode localization. Figure~\ref{fig: thkns_map}(d) and the residual errors between theoretical and estimated dispersion curves for each mode are discussed in Section~\ref{sec: k_variations}. This shows that as the wavenumbers increase, the uncertainty of our predictions also increases. Overall, mode 4 has the highest uncertainty. This observation is substantiated by Fig.~\ref{fig: gamma}, where higher modes tend to have overall higher $\gamma$. Figure~\ref{fig: dat_4_modes} rows (b) and (c) show that the optimal $\gamma$ associated with each mode in wavenumber domain (row (b)), shown at the top of each column in Fig.~\ref{fig: dat_4_modes} and in Fig.~\ref{fig: gamma}, increases as the corresponding peak amplitude area in frequency domain (row (c)) decreases. Since $\gamma$ is a measure of uncertainty around the exact physical solution, the less area associated to a mode, the higher $\gamma$.

\section{Conclusion}
\label{sec: conclusion}

We have introduced a new wave-informed regression algorithm. It provides an interpretable and robust wavefield analysis through wave mode decomposition. Wave-informed regression is also fast due to operating in the sparse spectral domain of the input wavefield. Our method reconstructs wavefield measurements, as shown in Section~\ref{sec: rec_performance} and Fig.~\ref{fig: dat_data_rec}, with an overall correlation coefficient of $0.68$. It achieves spatial separability between extracted modes, as demonstrated in Section~\ref{sec: k_variations} and Fig.~\ref{fig: dat_4_modes}, thus detecting defects and thickness variations. It also effectively estimates wavenumbers corresponding to those thicknesses, as shown in Section~\ref{sec: thkns_validation} and Fig.~\ref{fig: thkns_map}, with an overall residual error between theoretical and estimated wavenumbers of $48.12$~rad/m. Finally, it reconstructs the specimen thickness variations solely from wave mode estimates, with a correlation coefficient of $0.94$, as shown in Section~\ref{sec: thkns_validation} and Fig.~\ref{fig: thkns_map}. Wave-informed regression is, therefore, a training-free, interpretable and robust defect detection and characterization method.

Wave-informed regression in this work is limited to isotropic, planar, and solely spatial wave mode analysis. To expand this,  modifications to the physics operator design could be explored. Due to the setup of our framework, the physics operator can be modified without altering the overall architecture. Anisotropic wave propagation has been explored in references \cite{tran2025, subhash2025}, where a fractional form of the wave equation was leveraged to model anisotropic wavenumber domain filters. Further work on alternative definitions for the physics operator is ongoing. 

Overall, the method developed and outlined in this paper represents an efficient, robust, and fully interpretable defect detection and characterization tool. With further investigations into alternative physics operator designs, this framework can be adapted to address additional complexities and harness the power of linear machine learning architectures and physics to provide a powerful, novel defect detection and characterization method.

\bibliography{mybib}

\appendix 

\section{Discretized 2D Helmholtz Equation}
\label{appendix: A}
Using matrix notation, the two-dimensional Helmholtz equation in \eqref{eq: 2dHelmholtz} can be expressed as

\begin{align}
    \label{eq: helmholtz_discrete}
    \Delta^2 k^2\mathbf{Y} &=  -  \; \left( \mathbf{Y} \L_x  +  \L_y \mathbf{Y}   \right)
\end{align}
If we flatten the matrix and utilize properties of the vectorization function, we can represent this as 
\begin{align}
    \Delta^2 k^2\vect(\Y) = \Delta^2 k^2\y 
    &=  - \; \vect\left( \Y \L_x  +  \L_y \Y   \right) \nonumber \\
    &=  - \; \vect\left( \Y \L_x \right)  +  \vect\left( \L_y \Y   \right)\nonumber \\
    &=  - \; \left[ \L_x \otimes \I  +  \I \otimes \L_y \right] \y\nonumber \\
    \Delta^2 k^2\y &=  - \, (\L_x \oplus \L_y) \y \; .
\end{align}
Here, $\otimes$ represents a Kronecker multiplication operation and $\oplus$ represents a Kronecker sum operation.

\section{Diagonal Physics Operator}
\label{appendix: B}

We start with the discretized Helmholtz equation \ref{eq: helmholtz_discrete}, and apply the eigendecomposition of the Laplacian:

\begin{align*}
    \Delta^2 k^2\mathbf{Y} + \mathbf{L}_y \mathbf{Y} + \mathbf{Y} \mathbf{L}_x &= 0 \\
    \Delta^2 k^2\mathbf{Y} + \bm{\Gamma}_y \mathbf{K}_y \bm{\Gamma}_y^{-1} \mathbf{Y} + \mathbf{Y} \bm{\Gamma}_x \mathbf{K}_x \bm{\Gamma}_x^{-1}& = 0 \\
    \Delta^2 k^2\bm{\Gamma}_y \bm{\Gamma}_y^{-1} \mathbf{Y} \bm{\Gamma}_x  \bm{\Gamma}_x^{-1} + \\ 
    \bm{\Gamma}_y \mathbf{K}_y \bm{\Gamma_y}^{-1} \mathbf{Y} \bm{\Gamma}_x  \bm{\Gamma}_x^{-1} +
    \bm{\Gamma}_y \bm{\Gamma}_y^{-1} \mathbf{Y} \bm{\Gamma}_x \mathbf{K}_x \bm{\Gamma}_x^{-1} &= 0 
\end{align*}
We now apply the transformation $\widehat{\mathbf{Y}} = \bm{\Gamma}_y^{-1} \mathbf{Y} \bm{\Gamma}_x$ to recover:
\begin{align}
   \Delta^2 k^2\bm{\Gamma}_y \widehat{\mathbf{Y}} \bm{\Gamma}_x^{-1} + \bm{\Gamma}_y \mathbf{K}_y  \widehat{\mathbf{Y}} \bm{\Gamma}_x^{-1} + \bm{\Gamma}_y \widehat{\mathbf{Y}} \mathbf{\mathbf{K}}_x \bm{\Gamma}_x^{-1} = 0 \nonumber \\
  \bm{\Gamma}_y \left[\Delta^2 k^2\widehat{\mathbf{Y}} + \mathbf{K}_y \widehat{\mathbf{Y}}  +  \widehat{\mathbf{Y}} \mathbf{K}_x \right] \bm{\Gamma}_x^{-1} = 0 \nonumber \\
 \Delta^2 k^2\widehat{\mathbf{Y}} + \mathbf{K}_y\widehat{\mathbf{Y}}  +  \widehat{\mathbf{Y}} \mathbf{K}_x = 0
\end{align}
Finally, by applying the same vectorization technique discussed in \ref{appendix: A}, we recover the discrete, diagonal physics operator:
\begin{align}
    \widehat{\mathbf{P}}(k) &= \left[\Delta^2 k^2\I \ + \,  \left(  \mathbf{K}_x \oplus \mathbf{K}_y  \right)\right]
\end{align}

\section{Discrete Fourier Transform Operator as Laplacian Eigenvectors}
\label{appendix: C}

The Fourier transform of the second partial derivative of a function is
\begin{align*}
    \mathbf{\mathcal{F}}_t\left(\frac{\partial^2\mathbf{u}(x,y,t)}{\partial t^2}\right) &= - \, \omega^2 \widehat{\mathbf{u}}(x,y,\omega)
\end{align*}
In the discrete domain, we use the Discrete Fourier Transform (DFT) operator and the Laplacian operator as follows.
\begin{align}
    \mathbf{\mathcal{F}}_t \ \mathbf{L}_t \ \mathbf{u}(x,y,t) &= \bm{\Omega} \mathbf{\mathcal{F}}_t \bm{u}(x,y,\omega) \nonumber\\
    \mathbf{L}_t \ \mathbf{u}(x,y,t) &= \mathbf{\mathcal{F}}_t^{-1}\bm{\Omega} \mathbf{\mathcal{F}}_t \bm{u}(x,y,\omega)
\end{align}
Where now $\bm{\Omega}_{ii}= -\omega^2$ is a diagonal matrix of the harmonic frequencies, and the Discrete Fourier Transform operator assumes that the signal is periodic. This equation is the eigenvalue equation for the Laplacian operator $\mathbf{L}_t$, with eigenvalues $-\omega^2$, and eigenvector $\mathbf{\mathcal{F}}_t^{-1}$. The diagonalization of the Helmholtz equation discussed in \ref{appendix: B} shows that the eigenvector matrices are irrelevant to the final definition of the equation. Therefore, if the signal $\mathbf{u}(x,y,t)$ is periodic in $t$, the discrete Fourier transform of its second derivative with respect to $t$ is equivalent to the eigendecomposition of the Laplacian. By applying the transform over all dimensions, we arrive at the physics operator in \eqref{eq: helmholtz_laplace_physics3}.

\section{Solutions for $\y$, $k^2$, and $\gamma^2$}
\label{appendix: D}

\begin{align*}
\min_{\substack{\y_{1}, \ldots, \y_{M}, \k, \bm{\gamma}^2}} 
&\left\| {\x} - \sum_{m=1}^M{\y}_m  \right\|_2^2  \\
&+ \sum_{m=1}^M \left[ \gamma_m^2 \; \left\| {\y_m}  \right\|^2_2 + \frac{1}{\gamma_m^2}  \left\| {\mathbf{P}_m(k_m) \y_m}  \right\|^2_2 \right] \nonumber
\end{align*}
From \eqref{eq: cost_function}, we may expand all norms and recover:
\begin{align*}
    \Theta = \left(\x - \A\y_m \right)^H \left(\x - \A\y_m \right) + \gamma^2\y_m^H\y_m \\+ \frac{1}{\gamma^2}\y_m^H\mathbf{P}(k_m)^H\mathbf{P}(k_m)\y_m
\end{align*}
The closed-form solution for the reconstruction $\y$ follows:
\begin{align}
    \frac{\partial \Theta}{\partial \y_m} &= -2\A^H(\x-\A\y_m) + 2\gamma^2\y_m  \nonumber \\
    &\qquad + \frac{2}{\gamma^2}\mathbf{P}^H(k_m)\mathbf{P}(k_m)\y_m = 0 \nonumber \\
    \y_m &= \left[ \A^H\A + \gamma^2\I + \frac{1}{\gamma^2}\mathbf{P}(k_m)^H\mathbf{P}(k_m) \right]^{-1}\A^H\x
\end{align}
Now representing $\mathbf{P}(k) = (\mathbf{L}_x \oplus \mathbf{L}_y) +\I k^2$, we may find the single mode solution for the wavenumber $k_m^2$:

\begin{align}
    & \frac{\partial \Theta}{\partial k_m^2} \\
    &= \frac{\partial}{\partial k_m^2}\y_m^H\left[\left((\mathbf{L}_x \oplus \mathbf{L}_y)+\I k_m^2\right)^H\left((\mathbf{L}_x \oplus \mathbf{L}_y)+\I k_m^2\right)\right]\y_m\nonumber \\
    &= \y_m^H\left[(\mathbf{L}_x \oplus \mathbf{L}_y)^H + (\mathbf{L}_x \oplus \mathbf{L}_y) + 2\I k_m^2 \right]\y_m = 0 \nonumber \\
    k_m^2 &= \frac{\y_m^H\left[(\mathbf{L}_x \oplus \mathbf{L}_y)^H + (\mathbf{L}_x \oplus \mathbf{L}_y)\right]\y_m}{2 \y_m^H\y_m} \\
    &= \frac{\y_m^H\left[\mathbf{P}(0)^H + \mathbf{P}(0)\right]\y_m}{2 \y_m^H\y_m}
\end{align}
Finally, the single mode solution for the bandwidth $\gamma_m^2$:
\begin{align}
    \frac{\partial \Theta}{\partial \gamma_m^2} &= \frac{\partial}{\partial \gamma_m^2} \left[ \gamma_m^2\y_m^H\y_m + \frac{1}{\gamma_m^2}\y_m^H\mathbf{P}^H(k_m)\mathbf{P}(k_m)\y_m \right] = 0 \nonumber \\
    &= \y_m^H\y_m - \frac{1}{\gamma_m^4}\y_m^H\mathbf{P}^H(k_m)\mathbf{P}(k_m)\y_m = 0 \nonumber \\
    \gamma_m^2 &= \sqrt{\frac{\y_m^H\mathbf{P}^H(k_m)\mathbf{P}(k_m)\y_m}{\y_m^H\y_m}}
\end{align}

\end{document}